\documentstyle[twoside,epsfig]{hsproc}
\def\runauthor{R.M. Godbole, A. Grau, G. Pancheri and Y.N. Srivastava}
\def\shorttitle{Total  Cross-Sections}
\begin{document}
\begin{flushright}
IISc-CTS/09/01\\
LNF-01/009(P)\\
UG-FT-129/01\\
hep-ph/0104015
\end{flushright}

\begin{center}

{\large\bf  Total Cross Sections 
\footnote{Talk presented by GP at the International Meeting on 
Hadron Structure 2000, Stara Lesna, October 1-8,2000. }} \\

\vskip 25pt

{\bf                        Rohini M. Godbole } \\ 

{\footnotesize\rm 
                      Centre for Theoretical Studies, 
                    Indian Institute of Science, Bangalore 560 012, India. \\ 
                     E-mail: rohini@cts.iisc.ernet.in  } \\ 

\bigskip

{\bf                       A. Grau } \\ 

{\footnotesize\rm Departamento de F\'\i sica Te\'orica y del 
Cosmos, Universidad de Granada, Spain. \\
E-mail: igrau@ugr.es} \\
\bigskip

{\bf                       G. Pancheri } \\ 

{\footnotesize\rm 
                    Laboratori Nazionali di Frascati dell'INFN,  
                     Via E. Fermi 40, I 00044, Frascati, Italy. \\ 
                     E-mail: Giulia.Pancheri@lnf.infn.it  } \\
\bigskip

{\bf                                  Y.N. Srivastava}\\

{\footnotesize\rm 
               INFN, Physics  Department, U. of Perugia, Perugia, 
              Italy, \\ 
              Northeastern university, Boston,Mass.02115, USA,\\
              E-mail: srivastava@pg.infn.it } \\

\vskip 30pt

{\bf                             Abstract 
}

\end{center}

\begin{quotation}
\noindent
A unified approach to total cross-sections, based on the QCD contribution to 
the rise with energy, is presented for the processes $pp$, $p{\bar p}$, 
$\gamma p, \gamma \gamma, e^+e^- \rightarrow hadrons$. 
For proton processes, a discussion of the role 
played by soft gluon 
summation in   taming the fast rise due to mini-jets
is presented. For photon-photon processes, a comparison with other models 
indicates the need for precision measurements in both the
low and high energy region, likely only with measurements 
 at future Linear Colliders.
\end{quotation}
\newpage
\title{Total  Cross-Sections}
\author{R.M.Godbole\email{rohini@cts.iisc.ernet.in}}
{Center for Theoretical Studies, Indian Institute of Science, 
Bangalore 560 012, India}
\author{A. Grau\email{igrau@ugr.es}}{Departamento de F\'\i sica Te\'orica y del 
Cosmos, Universidad de Granada, Spain}
\author{Giulia Pancheri\email{pancheri@lnf.infn.it}}
{INFN Frascati National Laboratories, I00044 Frascati, Italy}
\author{Y.N. Srivastava\email{srivastava@pg.infn.it}}
{INFN, Physics  Department, U. of Perugia, Perugia, Italy\\
Northeastern university, Boston,Mass.02115, USA}
\abstract{
A unified approach to total cross-sections, based on the QCD contribution to 
the rise with energy, is presented for the processes $pp$, $p{\bar p}$, 
$\gamma p, \gamma \gamma, e^+e^- \rightarrow hadrons$. 
For proton processes, a discussion of the role 
played by soft gluon 
summation in   taming the fast rise due to mini-jets
is presented. For photon-photon processes, a comparison with other models 
indicates the need for precision measurements in both the
low and high energy region, likely only with measurements 
 at future Linear 
Colliders. }
\section{Introduction}
This paper describes our  approach towards a QCD description of total
cross-sections, examining both proton  and  photon
processses. In recent years, data have become available for 
a complete set of processes. In addition to proton-proton and
proton-antiproton, 
$\gamma p$ \cite{H1,ZEUS} and $\gamma \gamma$ \cite{L3,maria,OPAL}
total cross-sections are now known over a
sufficiently extended energy range to allow for the observation
of the rise with energy of all these  cross-sections and one can 
aim to obtain a unified description.   In what follows, we shall describe
a QCD based approach for the energy dependence of the total
cross-section, based on the Minijet model\cite{EMM,US2} concerning the
rise, and the use of soft gluon summation techniques, not only, to  tame 
this rise\cite{ff2} but also to obtain the early decrease. 
We start with purely hadronic cross-sections and then look in detail to
photon-photon processes.  For these, we discuss various models\cite{photon99}
and the precision needed for discrimination between  them. Some of these 
models  are based on the above mentioned QCD approach, others rely strictly 
on factorization\cite{aspen} while some  others are a mixture of 
both\cite{BSW,DL,SAS,GLMN,BKKS}.    

\section{A QCD Approach}
The task of describing the energy behaviour of total
cross-sections can be broken down into three parts:
\begin{itemize}
\item the rise
\item the initial decrease
\item the normalization 
\end{itemize}
It has been known for quite some time, now, that the rise\cite{rise}
can be obtained using the QCD calculable
contribution from the parton-parton cross-section, whose total
yield increases with energy\cite{minijets}. It is not easy however to properly
model this rise so as to reproduce correctly the slope,  and we
believe that
soft  gluon radiation, is the key to tame the
rise due to hard gluon radiation. The distinction between soft and
hard is so far arbitrary, but in this context we call soft gluons as those
which do  not undergo scattering against another
parton in the colliding hadron, hard gluons as those which
participate in the perturbative parton-parton scattering. This distinction
corresponds to saying that the soft gluons have wavelength too long to see the
content of the scattering area. It should be noted that the definition of the
scattering area is energy dependent, since at very high energy, our physical
picture of the scattering region is that 
of an expanding disk. We shall return to this point in later sections.

\begin{figure}[htb]
\begin{center}
\epsfig{file=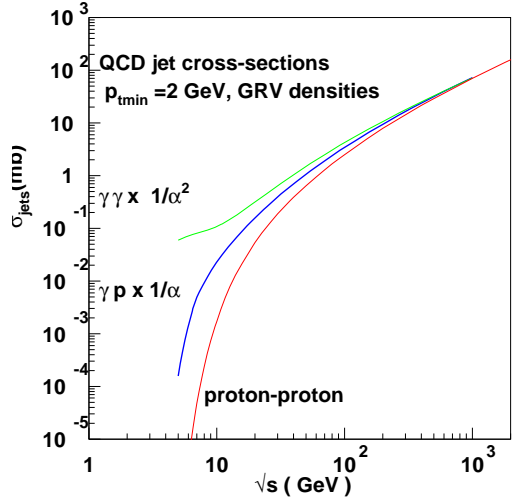,width=10cm}
\caption{Integrated jet cross-sections for $p_{\rm tmin}=2\ GeV$. }
\label{jets}
\label{mini}
\end{center}
\end{figure}

\begin{figure}[htb]
\begin{center}
\epsfig{file=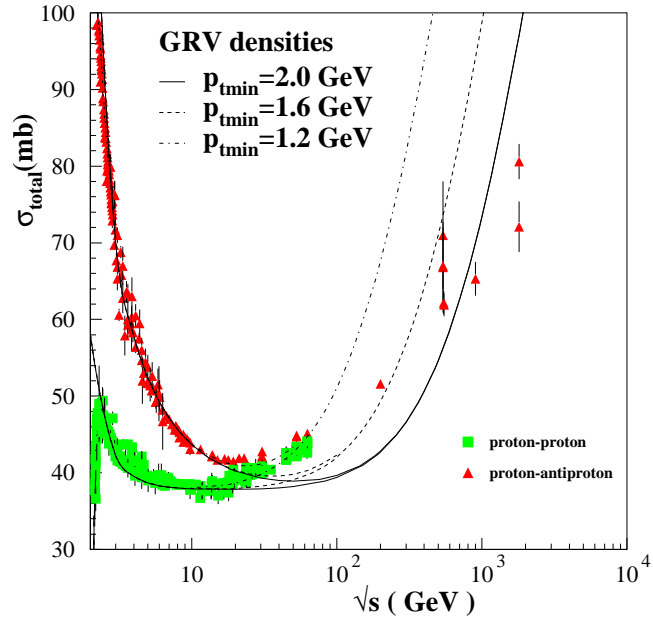,width=10cm}
\vspace{1cm}
\caption{Comparison between data\protect\cite{PDG} and predictions from the 
EMM (see text),
for different minimum jet transverse momentum.}
\label{EMM}
\end{center}
\end{figure}

To see how the jet cross-section can contribute to the rise, consider
the quantity
\begin{equation}
\sigma_{jet}(s;p_{tmin})=\int_{p_{tmin}} dp_t {{d\sigma^{jet}}\over{dp_t}}
=\sum_{i,j,k,l}\int f_{i/a}(x_1)dx_1 \int f_{j/b}(x_2)dx_2\int
{{d{\hat \sigma}(ij\rightarrow k l)}\over{dp_t}} dp_t 
\end{equation} 
where the sum goes over all parton types. This quantity
 is a function of the minimum transverse momentum
$p_{tmin}$ of the produced jets and
 can be calculated using the, phenomenologically 
determined, parton densities for protons
and
photons. We show in Fig.\ref{mini} the QCD jet cross-sections
with $p_{tmin}=2 \ GeV$ obtained using GRV\cite{GRV} densities for
the three processes $proton-proton, \gamma \ proton$ and $\gamma
\gamma$, normalized so as to be compared with each other. 
The
rise with energy for any fixed value of $p_{tmin}$
is clearly observed. Notice that the rise is stronger for
smaller values of $p_{tmin}$, whose value is a measure of the smallness
of the $x$-values probed in the collision. Thus different densities,
which correspond to different small-x behaviour, give
different results for the same $p_{tmin}$. In all cases however,
what one observes is that $\sigma_{jet}$ rises too fast to describe 
$\sigma_{tot}$. For a unitary description, the jet cross-sections
are embedded into the eikonal formalism\cite{eikminijets}, in which
\begin{equation}
 \label{eiktot}
\sigma^{\rm tot}_{pp(\bar p)}=2\int d^2{\vec b}
[1-e^{-\chi_I(b,s)}cos(\chi_R)]
\end{equation}
with $\chi=\chi_R+i\chi_I$, 
the eikonal function which contains both the
energy and the transverse momentum dependence of matter distribution
in the colliding particles, through the impact parameter distribution in 
b-space.
The physical picture described by the eikonal representation is that of
two colliding hadronic disks, with partons providing 
the basic scattering constituents,
distributed in the two hadrons according to an, {\it a priori}, unknown
matter distribution. Thus, schematically, at any given
c.m.  energy $\sqrt{s}$ and transverse distance,  
$\chi_I(b,s)$ should be obtained  by integrating the differential cross-section
over all subenergies and initial momenta of the colliding partons. In the
Eikonal Minijet Model (EMM) one approximates $\chi_R\approx 0$ and
calculates
$\chi_I$ through the average number of collisions, from the definition of
the inelastic cross-section. The simplest formulation, which
incorporates the mini-jet assumption that  it is the  jet cross-section
which drives the rise, is given by
\begin{equation}
2 \chi_I(b,s)\equiv n(b,s)=A(b) [ \sigma_{soft}+\sigma_{jet}]
\end{equation}
so as to separate the calculable part, $\sigma_{jet}$, from the
rest, to be parametrized.
The normalization depends both on $\sigma_{soft}$ and on  the
b-distribution. For the latter, the simplest hypothesis, is that
it is given by the Fourier transform of the  e.m.  form factors of the
colliding particles, i.e.
\begin{equation}
A_{ab}(b)\equiv A(b;k_a,k_b)=
{{1}\over{(2\pi)^2}}
\int d^2 {\vec q} e^{\vec q \cdot \vec b}
{\cal F}_a(q,k_a){\cal F}_b(q,k_b)
\end{equation}
where $k_i$ are the scale factors entering into the
form factors. 
The problem with such straightforward
formulation for protons is that presently used gluon  densities like GRV
have such an 
energy dependence that it is not possible, with the above scheme, and
without further approximations, to
simultaneously describe both the early rise and the high energy end.
For instance, it is possible to describe the early rise, which takes place
around $10\div 30\ GeV$ for proton-proton and
proton-antiproton scattering, using GRV densities and a $p_{tmin}\simeq  1
\ GeV$, but then the cross-sections start rising too rapidly, whereas a
$p_{tmin}\approx 2\ GeV$ can   reproduce the
Tevatron points\cite{CDF-D0,E710}, but it misses the early rise. This can be
seen in Fig.(\ref{EMM}).  
To cure this difficulty, a QCD model for the impact parameter distribution
of partons has been proposed\cite{ff2}. In this model the function $A(b,s)$
tames the rise at high energy induced by $\sigma_{jet}$. Technically this
happens because the proposed function, as the energy increases, is
more and more suppressed at large b. Physically this reflects the fact that
as the energy increases, the partons become more and more acollinear, due
to 
soft gluon emission, and the probability of
collisions is reduced. To obtain such effect, it was
proposed that the normalized function $A(b,s)$ is obtained as the Fourier
transform of the trasverse momentum distribution of
colliding parton pair. In the leading
order, the parton pair has zero transverse momentum as each
parton is emitted along the direction of its parent proton. But, at
high energy, this cannot be true, and the colliding valence quarks will have
a 
degree of acollinearity, which can be calculated using soft gluon resummation 
tecnniques. The distribution thus obtained, labelled
Bloch-Nordsieck distribution\cite{PRtransverse}, is the following
\begin{equation}
A_{BN}(b,s)=
{\cal F} [{\cal P}_{BN}]={{e^{-h(b,s)}}\over{\int d^2{\vec b} e^{-h(b,s)}}}
\label{abneq}
\end{equation}
with
\begin{equation}
h(b,s)={{8}\over{3\pi}}\int_0^{q_{max}}{{dk}
\over{k}}
\alpha_s(k^2)\ln({{q_{max}+\sqrt{q_{max}^2-k^2}}
\over{q_{max}-\sqrt{q_{max}^2-k^2}}})[1-J_0(kb)]
\label{hbeq}
\end{equation}
and $q_{max}$ is a slowly increasing function of energy, which
depends on the kinematics of the process\cite{greco}.

Such a  total cross-section formulation exhibits a scale dependence 
through the QCD coupling
constant $\alpha_s$, basically
\begin{itemize} 
\item 
through the well known and
clearly defined $p_t$-dependence in parton-parton collisions, which we 
take to be $\alpha_s(p_{tmin}^2)$ with $p_{tmin}\ge 1\div2\ GeV$, 
\item 
through the $k_t$ dependence of the initial colliding
partons, obtained from soft gluon emission. 
\end{itemize}
This latter dependence needs to be clarified further. The integral in
eq.(\ref{hbeq})  extends down to $k_t=0$ and one
needs to model the infrared behaviour of $\alpha_s$ in
order to carry through the quantitative application of this Bloch-Nordsieck
ansatz. It clearly follows from eqs.(\ref{abneq},\ref{hbeq}) that the more 
singular  $\alpha_s$ is
as $k_t\rightarrow 0$, the larger the function $h(b,s)$ is at
large $b$-values and the faster is the fall of the function $e^{-h(b,s)}$ 
with increasing $b$.
On the other hand, as the energy of the colliding particles increases,  
$q_{max}$ is larger causing the function $h(b,s)$ to be larger, 
producing a suppression at high energy, for large $b$.  The overall result 
is that the behaviour of $A(b,s)$  at large $b$ is determined by 
\begin{itemize}
\item higher $\sqrt{s}$ producing larger $q_{max}$ and more emission
\item singular behaviour of $\alpha_s$ in the infrared region producing also
many more soft gluons
\end{itemize}
The above considerations can be made quantitative, by introducing an average
over the parton densities and assuming an approximate factorization
between the transverse and the longitudinal degrees of
freedom, i.e.
\begin{equation}
n(b,s)\approx n_{soft} + n_{hard}\approx 
n_{soft}+A_{BN}(b,<q_{max}>) \sigma_{jet}
\end{equation}
 The following average expression for
$<q_{max}>$ was proposed in  \cite{ff2},
 \begin{equation}
\label{qmaxav} 
M\equiv <q_{max}(s)>={{\sqrt{s}}  \over{2}}{{ \sum_{i,j}\int
{{dx_1}\over{ x_1}} f_{i/a}(x_1)\int
{{dx_2}\over{x_2}}f_{j/b}(x_2)\sqrt{x_1x_2}
\int_{z_{min}}^1
 dz (1 - z)}
\over{\sum_{i,j}\int {dx_1\over x_1}
f_{i/a}(x_1)\int {{dx_2}\over{x_2}}f_{j/b}(x_2) \int_{z_{min}}^1 dz}}
\end{equation}
with $z_{min}=4p_{tmin}^2/(sx_1x_2)$ and $f_{i/a}$ the valence 
quark densities used in the jet cross-section calculation.

\begin{figure}[htb]
\begin{center}
\epsfig{file=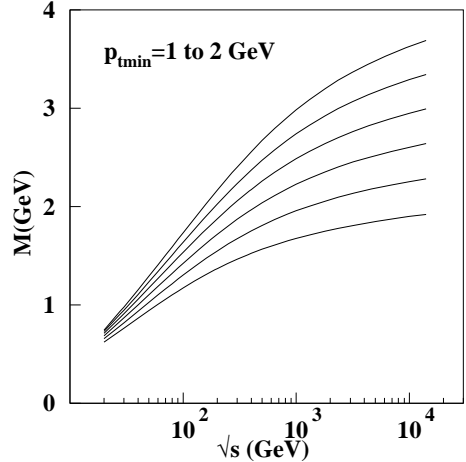,width=10cm}
\vspace{1cm}
\caption{The average maximum energy allowed to soft gluon emission in
jet production as a function of  $\sqrt{s}$ for various $p_{tmin}$
values and with GRV94 densities}
\label{qmax}
\end{center}
\end{figure}

M establishes the scale which, on the average, regulates soft gluon emission in the collisions, 
whereas $p_{tmin}$ provides the scale which
characterizes the onset of hard parton-parton scattering. 
For any parton parton subprocess characterized by a $p_{tmin}$  
of $1\div 2\ GeV$, M has a logarithmic  increase at 
reasonably low energy values and  an almost constant behaviour 
at high energy\cite{ff2}. 
In Fig.(\ref{qmax}) we plot the quantity M as a function of $\sqrt{s}$.

\begin{figure}[ht]
\begin{center}
\centerline{
\hspace{0.5in}
 \includegraphics*[height=2.80in,angle=90]{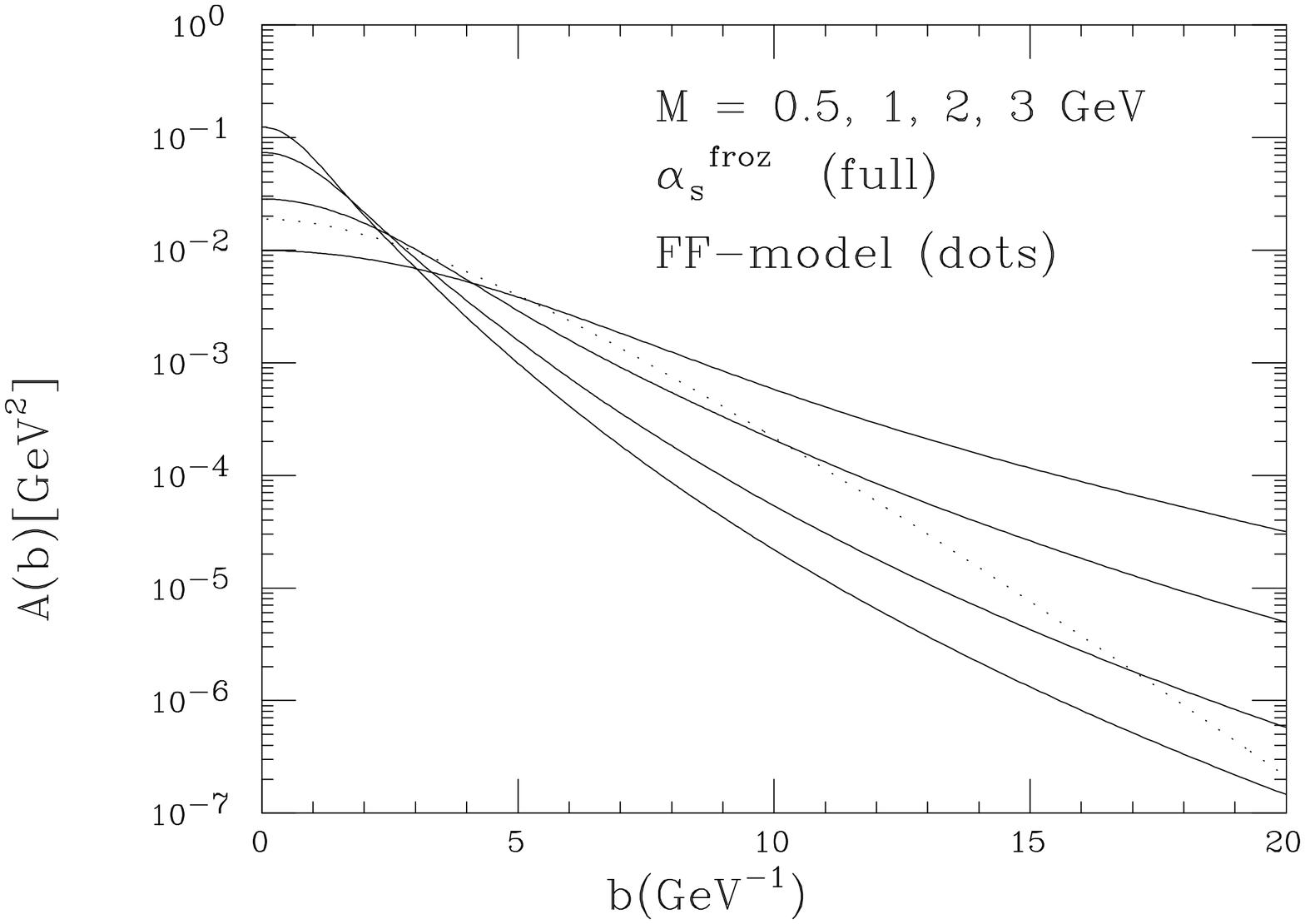}
 \includegraphics*[height=2.80in,angle=90]{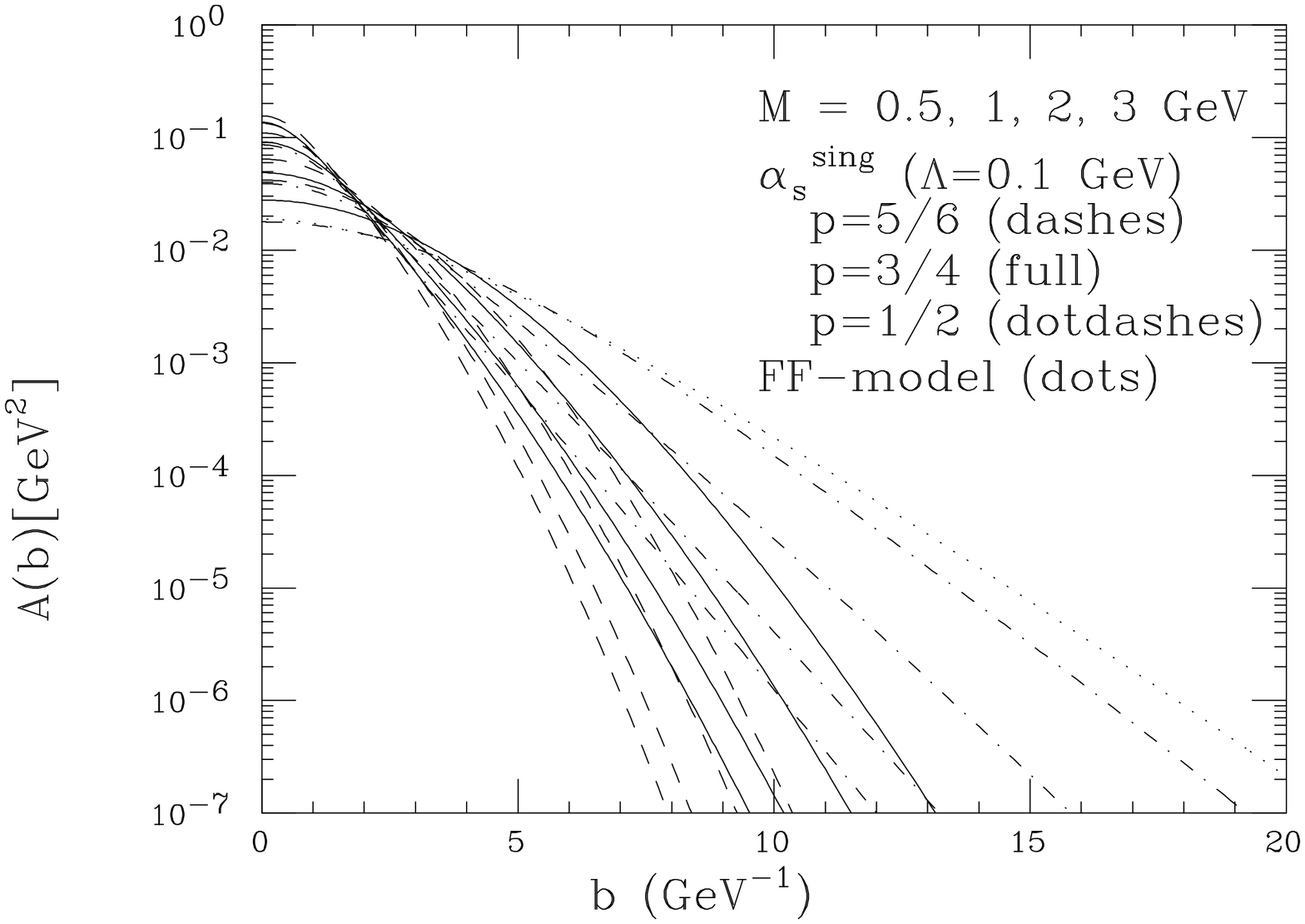}
}
\caption{Comparison of the A(b,s) for the Form factor model with that
in BN model,  for frozen $\alpha_s$ (left panel) and for singular $\alpha_s$, 
for various values of the parameter p (right panel).}
\label{ABN1}
\end{center}
\end{figure}

With the quantity $q_{max}$ thus determined, one can now calculate the
Bloch-Nordsieck
distribution function A(b,s), using different ans\"atze for
the functional expression for $\alpha_s$ in the $k_t$ going to zero limit.
It must be noticed that  what enters in all calculations is
not so much $\alpha_s$ as such, but rather its integral over the
infrared region. Thus in principle from a phenomenological point of view, 
even a
singular $\alpha_s$ can be used, provided it is integrable. Two models have 
been looked in detail so far, the frozen\cite{HALZEN,ALTARELLI} 
$\alpha_s$ model, in which
\begin{equation}
\alpha_s(k_t^2)={{b}\over{\log(a^2+k_t^2/\Lambda^2)}}
\end{equation}
and 
the singular\cite{richardson} $\alpha_s$ model, in which
\begin{equation}
\alpha_s(k_t^2)={{b'}\over{\log(1+(k_t^2/\Lambda^2)^{2p})}}\buildrel
{k_t\rightarrow 0} \over \longrightarrow {{1}\over{k_t^{2p}}}
\end{equation}
is singular, but integrable for $p< 1$.
One can now calculate the  overlap function $A_{BN}(b)$ for frozen
and singular $\alpha_s$ cases and compare it  with the Form Factor
model, as shown in Fig. \ref{ABN1}.

\begin{figure}[htb]
\begin{center}
\epsfig{file=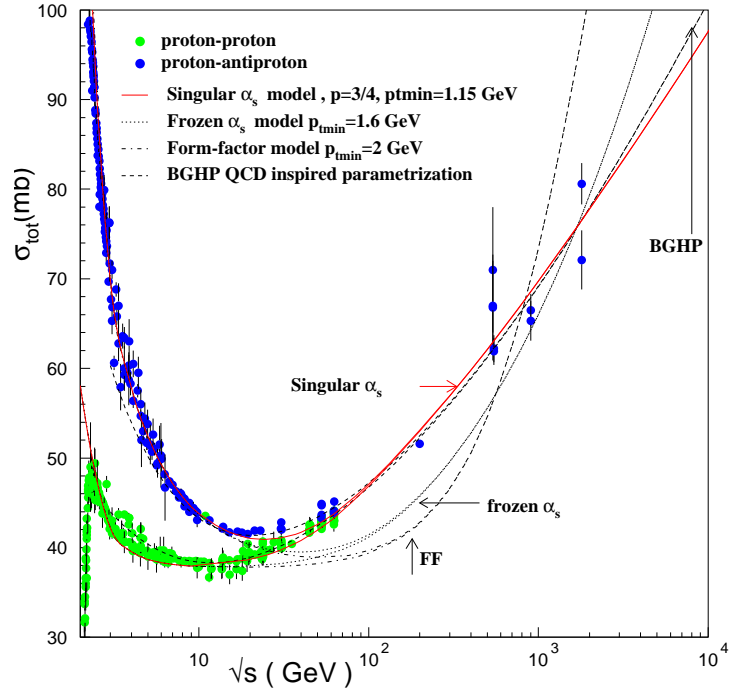,width=10cm}
\vskip 1cm
\caption{Total $p$-$p$ and ${\bar p}p$ cross-sections and comparison with various models}
\label{sigtotpp}
\end{center}
\end{figure}

The different behaviour of $A(b,s)$ in the large $b$ region,
changes the energy behaviour of the average number of collisions in this 
region, with the result that the cross-section indeed rises much less
rapidly when the singular $\alpha_s$ model is used. We 
compare  the three different models for the case
of  proton-proton and proton-antiproton scattering in Fig.(\ref{sigtotpp}),
where the early decrease and normalization have been described, in all three 
cases, using the Form factor model and
a parametrization of low energy data with 5 parameters. The rise on the
other hand is described using $\sigma_{jet}$ and the three different
b-distribution functions just described. Notice that, for the three models, we
have used different values of $p_{tmin}$ in the jet cross-section, since we
wanted to have  curves passing through the high energy points. We see that the
EMM model for protons using current  
parton densities like GRV does not reproduce  well the initial rise with
energy, and the same is also 
true for the frozen $\alpha_s$ model. For a comparison, we also show the
QCD inspired description, labelled BGHP,   used in the Aspen model
\cite{aspen} to predict 
photon cross-sections through factorization.
What this exercise shows is that the energy
behaviour of
the total cross-section is determined both by
soft gluon emission,  and
by hard gluons.  Work is in progress to determine
whether
soft gluon emission, which produces a decrease of
the cross-section with 
energy, plays a role also in the initial low
energy region, the so called
Regge region.

\section{Photon processes}
\begin{figure}[htb]
\begin{center}
\epsfig{file=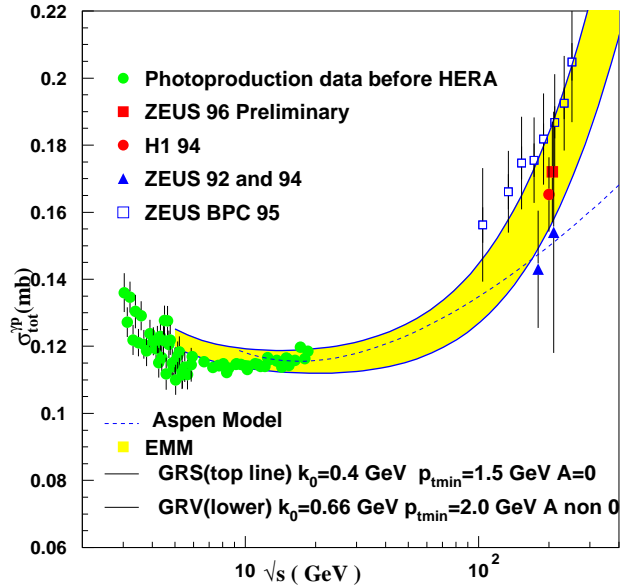,width=10cm}
\vspace{0.5cm}
\caption{Comparison between the eikonal minijet model predictions and data for 
total $\gamma p$ cross-section as well as BPC data extrapolated from 
DIS\protect\cite{DIS}. Predictions from \protect\cite{aspen} are also shown.}
\label{gp2k}
\end{center}
\end{figure}

\begin{figure}[htb]
\begin{center}
\epsfig{file=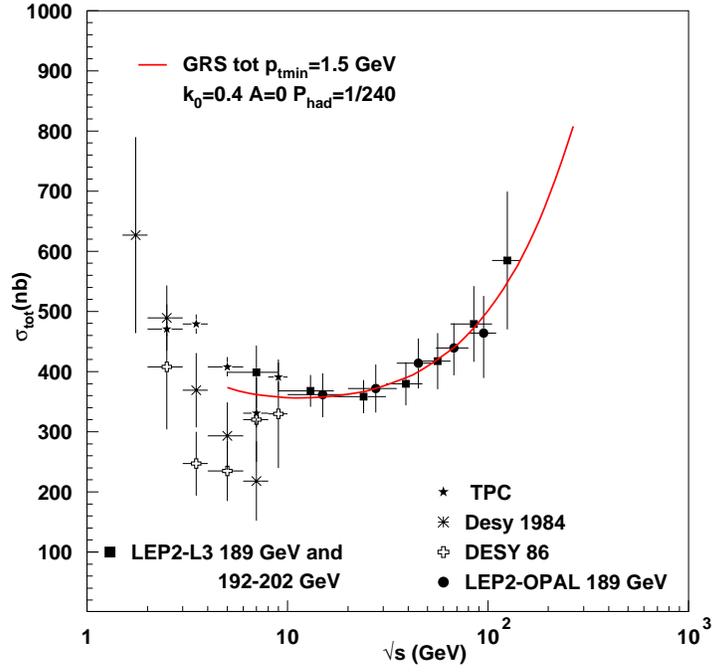,width=10cm}
\caption{The total photon-photon cross-section as described by the EMM (see text).}
\label{ggmaria}
\end{center}
\end{figure}

We now turn to discuss processes with photons, like photo-production or
$e^+e^-\rightarrow hadrons$, a process which, at high energy, is dominated by
$\gamma \gamma\rightarrow hadrons$. Again the main characteristics of the
photonic
cross-sections, are the overall normalization, the rise past 
$\sqrt{s}\approx
10 \ GeV$, and an initial decrease. For $\gamma \gamma$, the errors
on the overall normalization at low energy are so large that the initial
decrease is
hard to parametrized, not so for $\gamma p$. In this paper,  we
do not  address the question of the low energy behaviour, which is
simply obtained from the hadronic processes, through
scaling hypotheses.   For the normalization, one uses 
Vector Meson Dominance and Quark Parton Model ideas, by 
defining a quantity
$P_{had}=\sum_{V=\rho,\omega,\phi} {{4\pi\alpha}\over{f_V^2}}$ which
represents the probability that a photon exhibits a hadronic content.
The eiknoal formulation then has to be modified to take this into account
\cite{ladinsky} and is written as\cite{fletcher}
\begin{equation}
\sigma_{tot}=2 P_{had}\int d^2{\vec b}[1-e^{-\chi_I}(b,s)]
\end{equation}
with $2 \chi_I(b,s)=A(b)[\sigma_{soft}+\sigma_{jet}/P_{had}]$ with the 
notation of the previous section. In this approach, $P_{had}^{\gamma \gamma}
=(P_{had}^{\gamma p})^2 \approx (1/240)^2$, at 
$\sqrt{s_{\gamma \gamma}}
\approx 100\ GeV$. This factorization
ansatz seems to work reasonably well.
The soft part of the cross-section, which defines the
normalization, is obtained from
the number of quarks in the colliding particles,  through a simple 
scaling factor, i.e.
$\sigma_{soft}^{\gamma
p}={{2}\over{3}}\sigma_{soft}^{pp}$ and  $\sigma_{soft}^{\gamma
\gamma}={{2}\over{3}}\sigma_{soft}^{\gamma p}$. As for the rise, the
eikonal minijet
model of course uses jet cross-sections with the appropriate photon 
densities. We should mention that  there are at
present at least two other  models,  which obtain not only the
rise, but the entire cross-section from
the proton cross-section, using
factorization \cite{aspen,BSW} or a Regge-Pomeron
type behaviour\cite{DL,SAS}. We show in Fig.(\ref{gp2k})  old \cite{PDG}
 and recent
\cite{ZEUS-prelim} data for
$\gamma p$ total cross-section, together with BPC data extrapolated
\cite{bernd} from
DIS\cite{DIS},   compared with a band which represents the
predictions from the EMM model, using two different formulations\cite{photon2k}
with GRS\cite{GRS} and GRV\cite{GRV} type densities for the photonic partons.
This band is compatible with the predictions
for $\gamma \gamma$, for which we show in Fig. (\ref{ggmaria}),  our
favourite curve for the recent data from LEP\cite{maria}.  
The b-dependence in these figures is  obtained from an intrinsic transverse
momentum description~\cite{EMM}, using the experimentally measured transverse
momentum distribution of the photonic partons \cite{ZEUS-KT}. Work is 
in progress to apply to photonic processes
the Bloch-Nordsieck approach  described in the previous section. 
\section{ Precision needed to discriminate betwen models}
\begin{figure}[htb]
\begin{center}
\epsfig{file=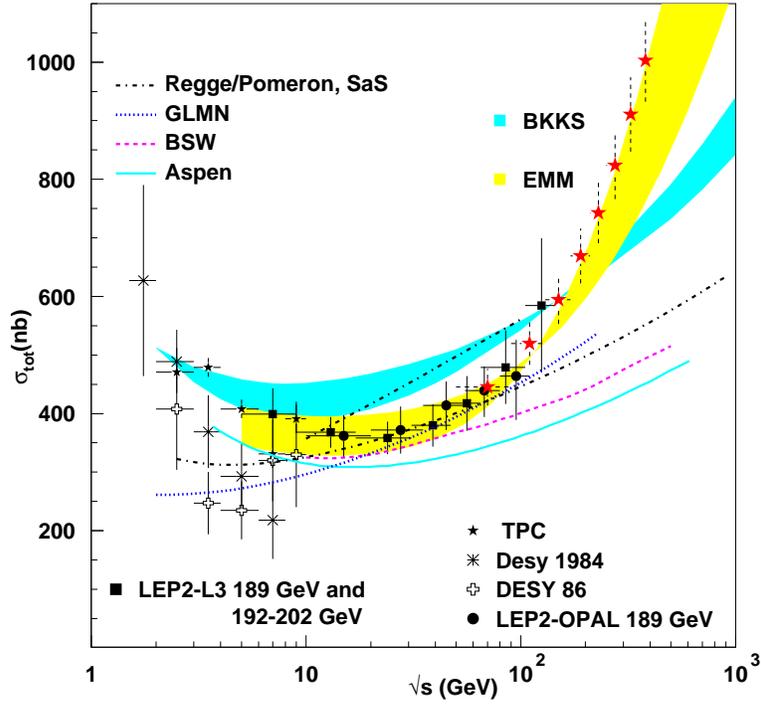,width=10cm}
\caption{Data and predictions from SaS\protect\cite{SAS}, 
GLMN\protect\cite{GLMN}, BSW\protect\cite{BSW},\protect\cite{aspen},
BKKS\protect\cite{BKKS} and the EMM model, described in the text}
\label{ggtotalbert}
\end{center}
\end{figure}

It is instructive to compare different models among each other, also
in view of
the proposed electron-positron Linear Collider which should reach very high
c.m.
energies, with $\sqrt{s_{\gamma \gamma}}$ potentially
as high as 500-700 GeV (if operated in the
photon collider mode). This comparison between various
model predictions  is shown in
Fig.(\ref{ggtotalbert}). The stars for large energy values correspond to
pseudo-data points, and illustrate a possible extrapolation of the
measured cross-section with realistic errors\cite{lcnote}.

In Table 1 we show total $\gamma \gamma$ cross-sections for  three
models of the "proton-is-like-the-photon' type. The last column shows 
the 1$\sigma$ level 
precision needed to discriminate between Aspen\cite{aspen}
and BSW\cite{BSW} models. The model labelled DL is obtained from
 Regge/Pomeron exchange with parameters from Donnachie and Landshoff\cite{DL}
 and
factorization at the residues. The difference between DL 
and either Aspen or BSW is bigger than between 
Aspen and BSW at each energy value.
\begin{table}
\begin{center}
\caption{ \label{tab:table1}Total $\gamma \gamma $ cross-sections and 
required precision for
models based on factorization}
\vspace{0.5cm}
\begin{tabular}{|c||c|c|c|c||} \hline 
$\sqrt{s_{\gamma \gamma}} (GeV)$ & Aspen &  BSW & DL & $1 \sigma$ \\ \hline
\hline
 20    & 309 nb & 330 nb & 379 nb &  7\%  \\ \hline
 50    & 330 nb & 368 nb & 430 nb &  11\%   \\ \hline
 100   & 362 nb & 401 nb & 477 nb &  10\%   \\  \hline
 200   & 404 nb & 441 nb & 531 nb &  9\%   \\  \hline
 500   & 474 nb & 515 nb & 612 nb &  8\%   \\  \hline
 700   & 503 nb & 543 nb & 645 nb &  8\%   \\ \hline
\end{tabular}
\end{center}
\end{table}

\begin{figure}[htb]
\begin{center}
\epsfig{file=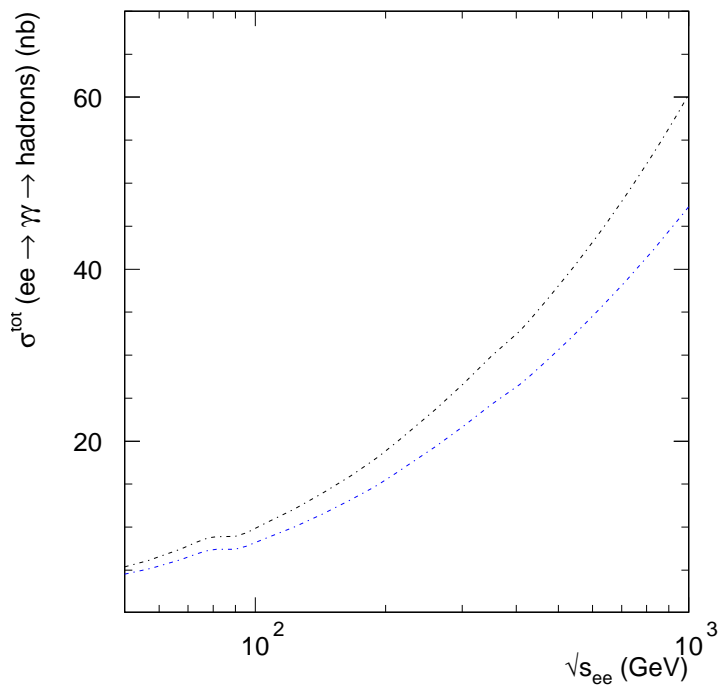,width=10cm}
\caption{Eikonal Minijet Model and Aspen Model,  in
 $e^+e^-\rightarrow hadrons$}
\label{folded}
  \end{center}
\end{figure}

Similar tables can be drawn for distinguishing among different
formulations of the Eikonal Minijet Model or between
the EMM and other QCD based models, like BKKS \cite{BKKS}. Part of the problem
in
comparing data with theoretical expectations
and hence make predictions for future machines, lies in the
fact that the total cross-section for photon processes
is difficult to measure, and theoretically 
difficult to define. A less uncertain
quantity is actually the $e^+e^-$ cross-section into hadrons, which
is the quantity from
which the $\gamma \gamma$ cross-sections are extracted. Thus it may be more 
appropriate to fold different model predictions for $\gamma \gamma$ 
cross-sections, with the photon distribution in the electrons taking into 
account the (anti)tagging of the electrons~\cite{DGV} and compare the
resultant uncertainties.
We show one such folding, using the Weisz\"acker Williams approximation,
with  two different models for $\sigma^{tot}_{\gamma\gamma}$, 
EMM with GRV densities and the Aspen model, in Fig.(\ref{folded}). 
We see that as a result of the folding, the difference between different
model predictions of a factor 2 or so get reduced to about $30 \%$. Given the
expected experimental errors at the future linear colliders it would be
possible to discriminate between different theoretical models for
$\gamma \gamma $ cross-sections even at a $e^+ e^-$ collider~\cite{lcnote}.

\section{Conclusion}
A QCD approach to the calculation of total cross-section based on the eikonal
formulation, shows that QCD can describe the rise for both, photon
and proton,  processes. For protons there exist extensive and very accurate
data, and it is seen that a
treatment of both hard and soft gluon emission needs to be used in order to
reproduce the rise  at intermediate and  high energies. Photon processes
are still characterized by large experimental errors and the Eikonal
Minijet Model, with only hard gluon scattering and fixed intrinsic
transverse momentum for the photon, appears adequate. However, there exist
models which are consistent with the current data within the experimental 
errors and which predict a much slower rise at higher energies. A
more complete treatment would be required for data with
much smaller experimental errors, likely only with 
measurements at the future Linear Colliders. 

\section{Ackowledgments}
We acknowledge support from the EEC, TMR contract 98-0169.

\end{document}